\documentclass[12pt,a4paper
]{article}
\usepackage{amsmath, amsthm,mathtools, amssymb,upgreek,slashed,hyperref}
\usepackage[utf8]{inputenc}
\usepackage{ifpdf}
\ifpdf
  \usepackage[pdftex]{graphicx}
  \usepackage{xcolor}
  \usepackage{epstopdf}
  \usepackage{bigints}
\else
  \usepackage[dvips]{graphicx}
\fi
\usepackage[paper=letterpaper,margin=0.9in]{geometry}
\usepackage{tikz}
\def\Bbb{\mathbb}

\def\SU{{\mathrm{\SU}}}

\def\OSp{{\textrm{OSp}}}
\def\SU{{\textrm{SU}}}
\def\SO{{\textrm{SO}}}
\def\USp{{\textrm{USp}}}
\def\Sp{{\textrm{Sp}}}
\def\Spin{{\textrm{Spin}}}
\def\Spin{{\textrm{Spin}}}
\def\Or{{\textrm{O}}}
\def\Un{{\textrm{U}}}

\def\U{{\mathrm U}}

\def\2{{\bf 2}}
\def\3{{\bf 3}}

\def\4{{\bf 4}}

\def\be{\begin{equation}}
\def\ee{\end{equation}}
 \def\\Sp{{\mathrm{\Sp}}}
 \def\\Spin{{\mathrm{\Spin}}}
 \def\\SU{{\mathrm{\SU}}}
 \def\\SO{{\mathrm{\SO}}}

\def\\Sp{\mathrm{\Sp}}

\def\R{{\Bbb{R}}}

\font\teneurm=eurm10 \font\seveneurm=eurm7 \font\fiveeurm=eurm5
\newfam\eurmfam
\textfont\eurmfam=\teneurm \scriptfont\eurmfam=\seveneurm
\scriptscriptfont\eurmfam=\fiveeurm

\font\teneusm=eusm10 \font\seveneusm=eusm7 \font\fiveeusm=eusm5
\newfam\eusmfam
\textfont\eusmfam=\teneusm \scriptfont\eusmfam=\seveneusm
\scriptscriptfont\eusmfam=\fiveeusm

\font\tencmmib=cmmib10 \skewchar\tencmmib='177
\font\sevencmmib=cmmib7 \skewchar\sevencmmib='177
\font\fivecmmib=cmmib5 \skewchar\fivecmmib='177
\newfam\cmmibfam
\textfont\cmmibfam=\tencmmib \scriptfont\cmmibfam=\sevencmmib
\scriptscriptfont\cmmibfam=\fivecmmib

\numberwithin{equation}{section}

\def\C{{\Bbb C}}

\def\\OSp{{\mathrm{\OSp}}}

\vspace{10pt}

\begin{document}
\begin{titlepage}
\begin{flushright}
\hfill{Imperial/TP/2021/mjd/3}\\

%\hfill{hep-th/yymmnnn}\\
\end{flushright}
\begin{center}

\vskip .5in %.3in 
\noindent

{\Large \bf{ The conformal brane-scan: an update }}
\bigskip\medskip

M. J. Duff\footnote{m.duff@imperial.ac.uk},\\
\bigskip\medskip

{\small

 Department of Physics, Imperial College London, Prince Consort Road, London SW7 2BZ
\vskip 3mm
\begin{center}
\&
\end{center}
Institute for Quantum Science and Engineering and Hagler Institute for Advanced Study, Texas A\&M University, College Station, TX, 77840, USA
}

\vskip .5cm %.3cm
\vskip .9cm %.6cm
     	{\bf Abstract }

\vskip .1in
\end{center}

\noindent

Generalizing the {\it The Membrane at the End of the Universe}, a 1987 paper {\it Supersingletons} by Blencowe and the author conjectured the existence of BPS $p$-brane configurations ($p=2, 3, 4 ,5$) and corresponding CFTs on the boundary of anti-de Sitter space with symmetries appearing in Nahm's classification of superconformal algebras: ${ \OSp(N|4)}~N=8, 4, 2, 1$; ${ \SU(2,2|N)}~N= 4, 2, 1$; 
${F^2(4)}$; ${\OSp(8^*|N)}, ~N=4, 2$. This correctly predicted the $D3$-brane with ${ \SU(2,2|4)}$ on $AdS_5 \times S^5$ and the $M5$-brane with ${ \OSp(8^*{}|4)}$ on $AdS_7 \times S^4$,  in addition to the known $M2$-brane with ${ \OSp(8|4)}$ on $AdS_4 \times S^7$.  However, finding non-singular AdS solutions matching the other symmetries was less straightforward.  Here we perform a literature search and confirm that all of the empty slots have now been filled, thanks to a number of extra ingredients including warped products and massive Type IIA. Orbifolds, orientifolds and S-folds also play a part providing examples not predicted: ${ \SU(2,2|3)}$, ${ \OSp(3|4)}$, ${ \OSp(5|4)}$ and ${ \OSp(6|4)}$ but not  ${ \OSp(7|4)}$.  We also examine the status of $p=(0,1)$ configurations.

\noindent
\vskip .5cm
\vskip .5cm

\vfill
\eject

\end{titlepage}

\setcounter{footnote}{0}
\setcounter{tocdepth}{2}

{\parskip = .2\baselineskip \tableofcontents}
\newpage
%\section{Introduction}

%\section{\bf Two brane-scans}
{\it Our mistake is not that we take our theories too seriously, but that we do not take them seriously enough.}

  Steven Weinberg

\section{Supersingletons}

\indent

The {\it Membrane at the End of the Universe} \cite{Duff:1987qa,Bergshoeff:1987dh,Blencowe:1987bn,Duff:1988st,Nicolai:1988ek,Bergshoeff:1988uc,Duff:1989ez,
Bergshoeff:1989av,Duff:1989aj,Claus:1998mw} was the name given to a supermembrane  \cite{Bergshoeff:1987cm} (later called the M2-brane) on  the $S^1 \times S^2$ boundary
 of $AdS_4 \times S^7$ described by a SCFT with symmetry 
\be
\OSp(8|4) \supset \SO(3,2) \times \SO(8) 
\ee
namely the $N=8$ singleton supermultiplet with 8 scalar and 8 spinors and $\SO(8)$ $R$ symmetry.  We recall that representations of $\SO(3,2)$ are denoted $D(E_0,s)$ where $E_0$ is the lowest energy eigenvalue which occurs and $s$ is the total angular momentum quantum number of the lowest energy state, analogous to the mass and spin of the Poincare group. However, Dirac's singletons $D(1/2,0)$ and $D(1,1/2)$ have no four-dimensional Poincare analogue \cite{Dirac:1963ta} and are best interpreted a residing on the three-dimensional boundary \cite{Fronsdal:1981gq,Nicolai:1984gb,Bergshoeff:1987dh}.
\begin{table}[h]
\small{
\begin{center}
\begin{tabular}{llllllllllllllllllllllllll}
d&G&H&&&Susy\\\\
6&$\OSp(8^{*}|N)$&$\SO^{*}(8) \times \USp(N)$&$N~ even$&&$8N$\\
5&$F^2{}(4)$&$ \SO(5,2) \times \SU(2)$&&&16\\
4&$\SU(2,2|N)$&$\SU(2,2) \times \Un(N)$&$N\neq 4$&&$8N$\\
~&$\SU(2,2|4)$&$\SU(2,2)\times \SU(4)$&&&32\\
3&$\OSp(N|4)$&$ \SO(N) \times \Sp(4, \R)$&&&$4N$\\
2&$G_{+} \times G_{-}$\\
1&$G_{\pm}=$\\
&$\OSp(N|2)$&$\Or(N) \times \SU(1,1)$&&&$2N$\\
& $\SU(N|1,1)$ & $\Un(N) \times \SU(1,1)$&$N \neq 2 $&&$4N$\\
&$\SU(2|1,1) $&$\SU(2) \times \SU(1,1)$&&&8 \\
&$\OSp(4^{*}|2N)$ &$  \SU(2) \times \U\Sp(2N) \times \SU(1,1)$&&&$8N$\\
&$G(3) $&$G_{2} \times \SU(1,1)$&&&14\\
&$F(4) $&$\Spin(7) \times \SU(1,1)$&&&16\\
&$D^{1}(2,1,\alpha) $&$\SU(2) \times \SU(2) \times \SU(1,1)$&&&8 \\
\end{tabular}
\end{center}}
\caption{Following \cite{Nahm:1977tg,VanProeyen:1999ni} we list the AdS supergroups in $d\le 6$ and their bosonic subgroups in the notation of \cite{Gunaydin:1998km}.}
\label{Nahm}
\end{table}
% with bosonic subgroups $\SO(p+1,2) \times \SO(D-p-1)$ describing $p$-branes on the boundary of  $AdS_{p+2} \times S^{D-p-2}$, as shown in Table \ref{horscan}. 

\begin{table}[h]
\begin{center}
\begin{tabular}{ccccccccccccccc}
&{\bf Supergroup}&{\bf Supermultiplet}&$B^-$&$V$&$\chi$&$\phi$&&$D$&\\\\
$AdS_3$&$\OSp(n|2) \times \OSp(8-n|2)$&$(n_+,n_-)=(n,8-n),d=2$~singleton&0&0&8&8&&10\\
&$\OSp(n|2) \times \OSp(4-n|2)$&$(n_+,n_-)=(n,4-n),d=2$~singleton&0&0&4&4&&6\\
&$\OSp(n|2) \times \OSp(2-n|2)$&$(n_+,n_-)=(n,2-n),d=2$~singleton&0&0&2&2&&4\\
&$\OSp(n|2) \times \OSp(1-n|2)$&$(n_+,n_-)=(n,1-n),d=2$~singleton&0&0&1&1&&3\\\\
$AdS_4$&$\OSp(8|4)$&$n=8,d=3$~singleton&0&0&8&8&&11&\\
&$\OSp(4|4)$&$n=4,d=3$~singleton&0&0&4&4&&7&\\
&$\OSp(2|4)$&$n=2,d=3$~singleton&0&0&2&2&&5&\\
&$\OSp(1|4)$&$n=1,d=3$~singleton&0&0&1&1&&4&\\\\
$AdS_5$&$\SU(2,2|2)$&$n=2,d=4$~doubleton&0&0&2&4&&8&\\
&$\SU(2,2|1)$&$n=1,d=4$~doubleton&0&0&1&2&&6&\\\\
&$\SU(2,2|4)$&$n=4,d=4$~doubleton&0&1&4&6&&10&\\
&$\SU(2,2|2)$&$n=2,d=4$~doubleton&0&1&2&2&&6&\\
&$\SU(2,2|1)$&$n=1,d=4$~doubleton&0&1&1&0&&4&\\\\
$AdS_6$&$F^2(4)$&$n=2,d=5$~doubleton&0&0&2&4&&9&\\\\
$AdS_7$&$\OSp(8^*|2)$&$(n_{+},n_{-})=(1,0),d=6$~tripleton&0&0&1&4&&10\\\\
&$\OSp(8^*|4)$&$(n_{+},n_{-})=(2,0),d=6$~tripleton&1&0&2&5&&11&\\
&$\OSp(8^*|2)$&$(n_{+},n_{-})=(1,0),d=6$~tripleton&1&0&1&1&&7&\\
\end{tabular}
\\
\end{center}
\medskip
\caption{Superconformal groups and their singleton, doubleton 
and tripleton representations. $B^-$, $V$, $\chi$, $\phi$ denote the number of chiral 2-forms, vector, spinors and scalars in each multiplet.  The spacetime dimension $D$ equals the worldvolume dimension $d$ plus the number of scalars.}
\label{fields}
\end{table}

Accordingly, in 1987
Blencowe and the author \cite{Blencowe:1987bn} conjectured the existence of  other BPS $p$-brane configurations with $p=(2,3,4,5)$ on the $S^1 \times S^p$ boundary of $AdS_{(p+2) }$ and corresponding CFTs with other symmetries appearing in Nahm's classification of superconformal algebras \cite{Nahm:1977tg}, listed in Table \ref{Nahm}. 

In each case the boundary CFT is described by the corresponding singleton (scalar), doubleton (scalar or vector) or tripleton (scalar or tensor) supermultiplet\footnote{Our nomenclature, based on the rank of $AdS_{p+2}$, is singleton $p=2$, doubleton $p=(2,3)$, tripleton $p=5$ and differs from that of G\"{u}naydin and Minic \cite{Gunaydin:1998km}.} as shown in Table \ref{fields}.  The number of dimensions transverse to the brane, $D-d$, equals the number of scalars in the supermultiplets.  None of these BPS brane CFTs is self-interacting. (For non-BPS see \cite{Seiberg:1999xz,Batrachenko:2002pu}).

\begin{table}
\small{
\begin{center}
\begin{tabular}{ccccccccccccccccc}
D$\uparrow$\\\\
%\hline\\
&&~{\bf SCALAR}\\\
11&.&&&${\bf OSp(8|4)}$&&&\\\
10&.&&$\OSp(n|2) \times \OSp(8-n|2)$&&&&$\OSp(8^*|2)$\\\
9&.&&&&&$F^2(4)$&\\\
8&.&&&&$\SU(2,2|2)$&&\\\
7&.&&&$\OSp(4|4)$&&&\\\
6&.&&$\OSp(n|2) \times \OSp(4-n|2)$&&$\SU(2,2|1)$\\\
5&.&&&$\OSp(2|4)$&&&\\\
4&.&&$\OSp(n|2) \times \OSp(2-n|2)$&$\OSp(1|4)$&&&\\\
3&.&&$\OSp(n|2) \times \OSp(1-n|2)$&~&&&\\\
2&.&&&&&&&&\\\
1&.&&\\\
0&.&.&.&.&.&.&\\\
&&~{\bf VECTOR}\\\
11&.&\\\
10&.&&&&${\bf SU(2,2|4)}$&&\\\
9&.&&&&&&\\\
8&.&&&&&&\\\
7&.&&&&&\\\
6&.&&&&$\SU(2,2|2)$&&\\\
5&.&\\
4&.&&&&$\SU(2,2|1)$&&\\\
3&.\\\
2&.&&&\\\
1&.&&\\
0&.&.&.&.&.&.&.\\\
&&~{\bf TENSOR}\\\
11&&&&&&&${\bf OSp(8^*|4)}$\\\
10&.&&&&&\\\
9&.&&&&&&\\\
8&.&&&&&&&\\\
7&.&&&&&&$\OSp(8^*|2)$\\\
6&.&&&&\\\
5&.&&&\\\
4&.&&&&&\\\
3&.&&&&&\\\
2&.&&&&&&&\\\
1&.&&&&&&\\\
0&.&.&.&.&.&.&.\\\
&0&1&2&3&4&5&6&d$\rightarrow$
\end{tabular}
\end{center}
\caption{The brane-scans of superconformal groups:
scalar supermultiplets: singletons ($p=1,2$), doubletons ($p=3,4)$ and tripletons ($p=5$);
vector supermultiplets: doubletons ($p=3$); 
tensor supermultiplets: tripletons ($p=5$). 
The M2-, D3- and M5-branes are in boldface.}
\label{tensor}}
\end{table}

A plot of spacetime dimension $D$ vs worldvolume dimension $d=p+1$, known as the {\it brane-scan}, is shown in Table \ref{tensor}.   This correctly predicted the $D3$-brane \cite{Horowitz:1991cd,Duff:1991pea,Duff:1992hu,Duff:1994an,Polchinski:1995mt,Polchinski:1996na} with ${\bf SU(2,2|4)}$ on $AdS_5 \times S^5$ and the $M5$-brane \cite{Gueven:1992hh,Duff:1992hu,Duff:1994an}  with ${\bf OSp(8^*|4)}$ on $AdS_7 \times S^4$,  in addition to the known $M2$-brane \cite{Bergshoeff:1987cm,Duff:1994an} with ${\bf OSp(8|4)}$ on $AdS_4 \times S^7$. 
The purpose of the present paper is to report  that all of the other slots have now been filled, thanks to a number of extra ingredients: warped products, massive Type IIA and Chern-Simons theories. Orbifolds, orientifolds and S-folds also play a part providing examples not predicted: ${ \SU(2,2|3)}$, ${ \OSp(3|4)}$, ${ \OSp(5|4)}$ and ${ \OSp(6|4)}$ but not  ${ \OSp(7|4)}$. We also examine the status of $p=(0,1)$ configurations. 
%\FloatBarrier

\section{The conformal brane-scan}

Comments:
\begin{itemize}
\item

The list in Table \ref{Nahm} is complete if one assumes that the Killing superalgebras of $AdS$ backgrounds are simple. However a more detailed investigation reveals that there may be some additional central generators in the Killing superlgebra for $AdS_3 $ and $AdS_5$ backgrounds \cite{Beck:2017wpm,Haupt:2018gap}

\item
The supersingleton lagrangian and transformation rules were also spelled out explicitly in \cite{Blencowe:1987bn}. This {\it conformal}  or (in later terminology) {\it near-horizon} brane-scan differs from the scan of Green-Schwarz type kappa-symmetric branes \cite{Achucarro:1987nc} which are not in general conformal and which, in any case, include only scalar supermultiplets. 
Further developments and elaborations on the brane-scan are summarized in \href{https://ncatlab.org/nlab/show/brane+scan}{Schreiber's n-lab} 
 and references  therein.
\item
In early 1988, Nicolai, Sezgin and Tanii \cite{Nicolai:1988ek} independently put forward the same generalization of the {\it Membrane at the End of the Universe} idea, spelling out the doubleton and tripleton lagrangian and transformation rules, in addition to the singleton.
However, by insisting on only scalar supermultiplets as in \cite{Achucarro:1987nc}  their list excluded the vector or tensor brane-scans of Table \ref{tensor}. In this case, as they 
point out, the spheres are just the parallelizable ones $S^1$, $S^3$ and $S^7$.

\item
 The two factors appearing in the $p=1$ case, $G_+ \times G_-$, are simply a reflection of the ability of strings to have left and right movers on the worldsheet \cite{Gunaydin:1986cs}.  
In this case, there are many candidate supergroups as shown in Table \ref{Nahm}, so for $p=0,1$ we did not attempt a complete list of which of these would eventually be realized. In  \cite{Blencowe:1987bn}, we focused on Type IIA, Type IIB and heterotic strings
with ${  \OSp(n|2)}_c \times { \OSp(8-n|2)}_s$, $ { \OSp(n|2)}_c \times { \OSp(8-n|2)}_c$ and $ { \OSp(n|2)}_c\times { \Sp(2, \R)}$, respectively, since the singleton CFTs (but not the supergravity $AdS_3$ solutions) had already been identified \cite{Gunaydin:1986cs}.  For concreteness the Type IIA case appears on the scan of Table \ref{tensor}. 

\item
Even for $p \geq 2$ not all  of the conformal algebras listed in Table \ref{Nahm} appear in the scan. For example, since none of our CFTs is self-interacting, we restricted  \cite{Blencowe:1987bn} ${ \SU(2,2|N)} $ to $N=1, 2, 4$ since perturbatively $N=3$ implies $N=4$. But we now know there are nonperturbative interacting CFTs with just $N=3$ \cite{Ferrara:1998zt,Aharony:2015oyb,Garcia-Etxebarria:2015wns,Aharony:2016kai,Borsten:2018jjm}.
We also focussed on $ N=1, 2, 4, 8$ in ${\OSp(N|4)}$ since they corresponded to the division algebra $\R, \C, \mathbb{H}, \mathbb{O}$ interpretation  of the four diagonal lines in the scalar branescan of Table \ref{tensor}. The $N=3,5,6,7$ cases are discussed in Section \ref{missing}.
\end{itemize}

\section{Significance of the brane-scan}

The significance of the $M2$, $D3$ and $M5$ and indeed the other  configurations on the brane-scan became clearer thanks to four major developments:

\begin{itemize}
 \item Branes as solitons 
 
The realization that string theory admits p-branes as solitons \cite{Townsend:1987yy,Strominger:1990et,Duff:1991sz,Duff:1991pe,Duff:1991pea,Horowitz:1991cd,Callan:1991at,Gibbons:1993xt,Duff:1994an}

 \item M-theory
 
  The realization that the Type IIA superstring in $D=10$ could be interpreted \cite{Duff:1987bx} as a wrapped supermembrane in $D=11$ \cite{Bergshoeff:1987cm}.  
The membrane is a 1/2 BPS solution of $D=11$ supergravity \cite{Duff:1990xz}, whose spacetime approaches Minkowski space far away from the brane but  $AdS_4 \times S^7$ close to the brane, jumping to the full ${\OSp(8|4)}$ in the limit \cite{Duff:1994fg}.  Regarded as an extremal black-brane, this limit was also called the near-horizon limit. Moreover multi-brane solutions could be obtained by stacking $N$ branes on top of one another \cite{Duff:1990xz}, yielding  quantized 4-form flux.  So $AdS_4 \times S^7$ could equally well be regarded as the large $N$ limit. A similar story applied to its magnetic dual fivebrane \cite{Gueven:1992hh} as a solution of $D=11$ supergravity. Moreover, the five string theories were merely different corners of an overarching M-theory \cite{Hull:1994ys,Witten:1995ex,Duff:1996aw} with $D=11$ supergravity as its low-energy limit. The membrane and fivebrane  were accordingly renamed M2 and M5.

\item D-branes

The realization that p-branes carrying RR charge, with a closed-string interpretation as solitons, admitted an alternative open string interpretation as Dirichlet-branes, surfaces of dimension $p$ on which open strings can end \cite{Polchinski:1996na}.  In particular the self-dual 3-brane, a solution of Type IIB supergravity with $AdS_5 \times S^5$ and ${ \SU(2,2|4)}$ in the large N limit, was reinterpreted as a D3-brane and renamed accordingly. 

\item AdS/CFT

The AdS/CFT  conjecture \cite{Maldacena:1997re,Gubser:1998bc,Witten:1998qj} proposes that large N limits of certain conformal field theories in d dimensions can be described in terms of supergravity (and string theory) on the product of d+1-dimensional AdS space with a compact manifold. Another vital ingredient, missing in the early days, was the non-abelian nature of the symmetries that appear when we stack N branes on top of one another \cite{Witten:1995im}. Examples include $N=4$ Yang-Mills in $D=4$ from $AdS_5 \times S^5$ and ABJM theory \cite{Aharony:2008ug} from $AdS_4 \times S^7/Z_n$.

\end{itemize}
    
\section{The missing ingredients $p\geq2$}
\label{missing}
Notwithstanding the success with $M2$, $D3$ and $M5$, for quite some time the status of the other slots on the brane-scans remained obscure \footnote{In  \cite{Duff:2008pa} we entertained the idea that they might arise from classical branes whose symmetry is enhanced when $\alpha'$ corrections are taken into account, but this did not pan out.}.
Here we perform a literature search and confirm that  all of the empty slots have now been filled, largely thanks to warped products, massive Type IIA, and Chern Simons theories as shown below

\begin{itemize}
\item{d=6}
$\OSp(8^*|N)$ $N=4,2$; \cite{Gueven:2003uw,Edelstein:2010sx,Apruzzi:2013yva, Gaiotto:2014lca, Cremonesi:2015bld,Cordova:2016emh,Nunez:2018ags} 

\item{d=5} 
$F^2(4)$ \cite{Brandhuber:1999np,Bergman:2012kr,Lozano:2012au,DHoker:2016ujz,DHoker:2016ysh,Cordova:2016emh,DHoker:2017zwj,
Corbino:2017tfl,Lozano:2018pcp}

\item{d=4}
$ \SU(2,2|N)$ $N= 4, 3, 2, 1$; 
\cite{Horowitz:1991cd,Gaiotto:2009gz,ReidEdwards:2010qs, Aharony:2012tz,Cordova:2016emh,Nunez:2019gbg,Ferrara:1998zt,Aharony:2015oyb,Garcia-Etxebarria:2015wns,Aharony:2016kai,Borsten:2018jjm}.

\item{d=3}
$\OSp(N|4)$  $N=8,6,5,4,3,2,1$ \cite{Duff:1990xz,Schwarz:2004yj,DHoker:2007zhm, DHoker:2008lup,Aharony:2008ug,Chiodaroli:2011nr, Assel:2011xz,Cordova:2016emh,Haupt:2017bnj,Marchesano:2020qvg,Bandres:2008ry}.

\end{itemize}
Comments
\begin{itemize}
\item
We have included $N=3$ in the $d=4$ case and $N=3,5,6$ in the $d=3$ case, which, as already mentioned ,were not predicted in  \cite{Blencowe:1987bn}. $N=6$ appears in ABJM \cite{Aharony:2008ug}. and its  ${\OSp(6|4)}$  symmetry in \cite{ Bandres:2008ry}.  A useful reference on the absence of $N=7$ is \cite{Cordova:2016emh}.

\item
There are no $AdS_7$ solutions in Types IIA and IIB. In M all are locally isometric to $AdS_7 \times S^4$.
\item
There are no maximally supersymmetric $AdS_6$ backgrounds in M, IIA or IIB. There are no half BPS (16 supersymmetries) $AdS_6$ backgrounds in M and IIA with compact internal space.
\item
There are no such $AdS_5$ solutions that preserve $>16$ supersymmetries in IIA and D=11
In IIB, all supersymmetric solutions are locally isometric to $AdS_5\times S^5$. This means
that all backgrounds preserving 24 supersymmetries in IIB are locally   $AdS_5\times S^5.$
\item

There are  no $ >16$ $AdS_4$ supersymmetric solutions in IIA and IIB.
In D=11 all  $>16$ supersymmetric solutions are locally isometric to $AdS_4\times S^7$.
This means that all solutions with 20, 24, 28 are locally  $AdS_4\times S^7.$
\end{itemize}

\section{$p=0,1$}
\begin{itemize}

\item{d=2}
\cite{Edelstein:2010sx,Lozano:2016kum,Lozano:2016wrs,Couzens:2017way,Lozano:2017ole,Itsios:2017cew, 
Haupt:2017bnj,Lozano:2019emq,Lozano:2019jza,Lozano:2019zvg,Lozano:2019ywa,Lozano:2020bxo,Faedo:2020nol,Faedo:2020lyw,Dibitetto:2020bsh,Macpherson:2021lbr}

\item{d=1} 
 \cite{Dibitetto:2018gtk, Gauntlett:2006ns, Kim:2013xza, Chiodaroli:2009yw, Chiodaroli:2009xh,Corbino:2018fwb,Corbino:2020lzq,Dibitetto:2019nyz,Lozano:2020txg,Lozano:2020sae,DHoker:2007mci,Bachas:2013vza}

\end{itemize}

Comment
\begin{itemize}
\item
Not all of the algebras in Nahm's list correspond to known solutions and indeed there may be some for which no solutions exist. 
A thorough and up-to-date summary maybe found in \cite{Macpherson:2021lbr}.

\end{itemize}

\section{Conclusion} 
Thus not only the M2, D3 and M5 but all of the $p$-brane configurations on the $S^1 \times S^p$ boundary of $AdS_{(p+1) }$ with $p=(5,4,3,2,1)$ mentioned explicitly in the 1987 paper as shown in Table  3 have now been discovered: ${ \OSp(N|4)}~N=8, 4, 2, 1$; ${\bf \SU(2,2|N)}~N= 4, 2, 1$; 
${ F^2(4)}$; ${ \OSp(8^*|N)}, ~N=4, 2$,  as have most of the $(p=0,1)$ in Nahm's list not mentioned explicitly. Orbifolds, orientifolds and S-folds also play a part providing examples not predicted: ${ \SU(2,2|3)}$, ${ \OSp(3|4)}$, ${ \OSp(5|4)}$ and ${\OSp(6|4)}$ but not  ${ \OSp(7|4)}$.  
To be fair, if our colleagues did not take our vector and tensor brane-scans seriously in 1987, it may be because, in the Weinberg sense, we did not take them seriously enough ourselves. 
\section*{Acknowledgements}
 
Correspondence with Fernando Alday, Connor Behan, Leron Borsten, Eric D'Hoker, George Papadopoulos, Yolanda Lozano and  Alessandro Tomasiello is greatly appreciated. I am grateful to Marlan Scully for his hospitality in the Institute for Quantum Science and Engineering, Texas A\&M University, and to the Hagler Institute for Advanced Study at Texas A\&M for a Faculty Fellowship. 
This work was supported in part by the STFC under rolling grant ST/P000762/1. 
\bibliography{Bibliography}
%\bibliography{references}
%\bibliographystyle{bibliography}   
\end{document}